\documentclass[aps,prl,twocolumn,showpacs,superscriptaddress,groupedaddress,longbibliography,citeautoscript]{revtex4-1}
\usepackage[T1]{fontenc}
\usepackage[latin9]{inputenc}
\usepackage{amsmath}
\usepackage{amssymb}
\usepackage{graphicx}
\usepackage[unicode=true,pdfusetitle,
 bookmarks=false,
 breaklinks=false,pdfborder={0 0 0},backref=false,colorlinks=false]
 {hyperref}

\makeatletter

\DeclareRobustCommand{\greektext}{%
  \fontencoding{LGR}\selectfont\def\encodingdefault{LGR}}
\DeclareRobustCommand{\textgreek}[1]{\leavevmode{\greektext #1}}
\DeclareFontEncoding{LGR}{}{}
\DeclareTextSymbol{\~}{LGR}{126}
\providecommand{\tabularnewline}{\\}

 \@ifundefined{textcolor}{}
 {%
   \definecolor{BLACK}{gray}{0}
   \definecolor{WHITE}{gray}{1}
   \definecolor{RED}{rgb}{1,0,0}
   \definecolor{GREEN}{rgb}{0,1,0}
   \definecolor{BLUE}{rgb}{0,0,1}
   \definecolor{CYAN}{cmyk}{1,0,0,0}
   \definecolor{MAGENTA}{cmyk}{0,1,0,0}
   \definecolor{YELLOW}{cmyk}{0,0,1,0}
 }

\usepackage{bbold}\usepackage{diagbox}\usepackage{makecell}\usepackage{bbm}

\setcitestyle{super}

\makeatother

\begin{document}

\title{Complementarity Reveals Bound Entanglement of Two Twisted Photons}

\author{Beatrix C. Hiesmayr}

\affiliation{Institute of Theoretical Physics and Astrophysics, Masaryk University,
Kotlá\v{r}aská 2, 61137 Brno, Czech Republic}

\affiliation{University of Vienna, Faculty of Physics, Boltzmanngasse 5, A-1090
Vienna, Austria}

\author{Wolfgang Löffler}

\affiliation{Leiden University, Quantum Optics \& Quantum Information, PO Box
9500, 2300 RA Leiden, Netherlands}
\begin{abstract}
We witness for the first time the generation of bound entanglement
of two photon qutrits, whose existence has been predicted by the Horodecki
family in 1998. Detection of these heavily mixed entangled states,
from which no pure entanglement can be distilled, is possible using
a key concept of Nature: complementarity. This captures one of the
most counterintuitive differences between a classical and quantum
world, for instance, the well-known wave-particle duality is just
an example of complementary observables. Our protocol uses \emph{maximum}
complementarity between observables: the knowledge about the result
of one of them precludes any knowledge about the result of the other.
It enables ample detection of entanglement in arbitrary high-dimensional
systems, including the most challenging case, the detection of bound
entanglement. For this we manipulate ``twisted'' twin photons in
their orbital angular momentum degrees of freedom. Our experimentally
demonstrated ``\emph{maximum complementarity protocol}'' is very
general and applies to all dimensions and arbitrary number of particles,
thus enables simple entanglement testing in high-dimensional quantum
information and opens up the quest of understanding the meaning of
this type of entanglement in Nature.
\end{abstract}
\maketitle
The boundary between a classical and quantum world is the domain of
mixed quantum states. Quantum entanglement of mixed states is much
more complex than that of pure states and subject of intense research.
In 1998 it was found that entanglement of mixed states is not ``flat'',
but has an intriguing structure \cite{horodecki1998}: Entangled quantum
states can be classified into two distinct types, those that can and
those that cannot be distilled into pure entangled states using statistical
local operations and classical communication. Undistillable entangled
states are called bound entangled. Bound entanglement can occur in
bipartite systems with dimensions larger than $d=2$ or in the case
of more particles. These two different types of bound entanglement
differ considerably. The case of more than $2$ entangled qubits has
recently been discovered experimentally in photonic multipartite systems
\cite{amselem2009,lavoie2010b,kaneda2012}, trapped ions \cite{barreiro2010},
NMR \cite{kampermann2010}; further, there is one study in a continuous-variable
context \cite{diguglielmo2011}. These results on multipartite qubit
bound entanglement became already interesting for certain quantum
information tasks such as steering \cite{brunner2012b} and reduction
of the communication complexity \cite{epping2012}. We investigate
here the case of bound entanglement of \emph{only two} photonic qutrits
($d=3$) using the orbital angular momentum degree of freedom of light;
this is the simplest case of bound entanglement and complications
such as occurring for multipartite systems \cite{lavoie2010} do not
occur. The orbital angular momentum of photons \cite{allen1992,molinaterriza2007}
is used in a number of studies to obtain high-dimensional bipartite
entangled qudits \cite{mair2001,dada2011,salakhutdinov2012,shalm2012};
apart from fundamental questions such as those discussed here, the
system shows promise in quantum cryptography \cite{bechmannpasquinucci2000a}
due to improved security, increased resistance against noise, and
higher bit rates \cite{kaszlikowski2000,fujiwara2003,bechmannpasquinucci2000}
compared to qubits. In particular for such higher-dimensional qudit
systems, it is well known that it is extremely hard to check whether
a given state, especially if it is mixed -- the daily situation in
a laboratory -- is separable or entangled. There are several operational
criteria available to detect entanglement, where the most famous and
powerful one is the Peres-Horodecki criterion, which is based on the
partial transposition in one subsystem \cite{peres1996,horodecki1996}.
After partial transposition, either all eigenvalues remain positive
(PPT), or in the other case, if at least one eigenvalue is found to
be negative, the state is entangled. The transposition of a state
is synonymous to time reversal \cite{sanpera1998}, so PPT tests if
a state where time's arrow is reversed for one of its partitions is
still a physical state: obviously, if the state is separable, this
must be the case. For two qubits, the reverse is also true and PPT
is a sufficient criterion for separability. This argument fails already
for two qutrits, where it turns out that bound entangled qutrits are
PPT. This brings us back to distillability: it was found that it is
actually the positivity under partial transposition, which makes bound
entangled states undistillable \cite{horodecki1998}. In the experiment
described here, in a first step, we produce photon states that are
bound entangled in their orbital angular momentum degree of freedom
and verify the positivity under partial transposition via state tomography.
In a second step we apply the maximum complementarity protocol using
sets of observables, for which their eigenbases are mutually unbiased,
to witness directly the inseparability of two bound entangled qutrits.

\begin{figure}
\includegraphics[width=0.99\columnwidth]{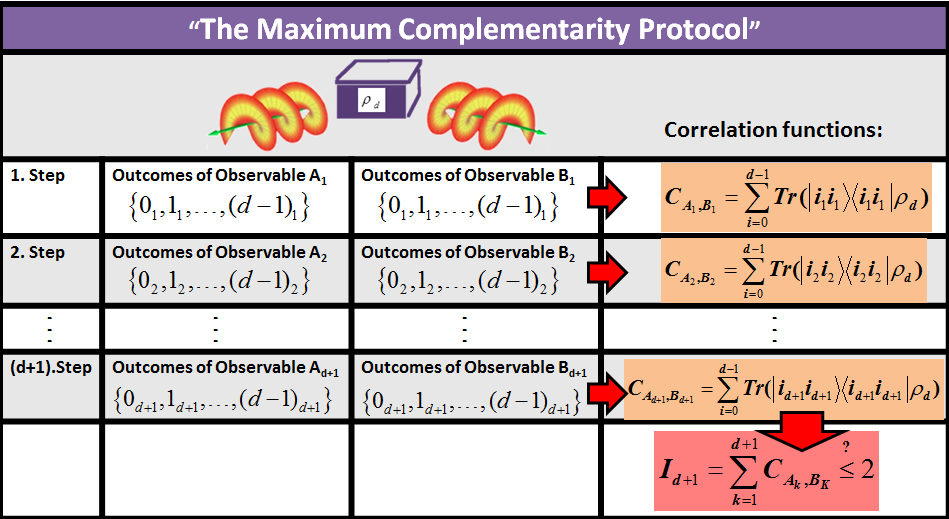} \caption{\label{fig:protocol} This figure illustrates the steps involved in
the complementarity protocol for entanglement detection of bipartite
qudits: Alice and Bob determine the experimental probability (coincidence
counts) for each detection state, and that for all (mutually unbiased)
bases. It is then simply a matter of summing up these probabilities,
if the sum is $>2$, the state is entangled. Note that both Alice
and Bob are allowed to relabel their measurement outcomes per basis,
in order to optimize the detection ability.}
\end{figure}

\textbf{The Maximum Complementarity Protocol:} Consider the following
scenario of a source producing two-qudit states $\rho\in\mathbb{C}^{d\times d}$,
namely quantum states with $d$ degrees of freedom per qudit. Both
experimenters, Alice and Bob, can choose among $k$ different observables.
What is the best strategy for Alice and Bob to detect the inseparability?
The most striking difference between entanglement and separability
are revealed by correlations in different basis choices. A fully correlated
system is a physical system for which we can predict with certainty
the outcome of a second measurement when we know the outcome of a
first measurement, opposite to the other extreme case when knowledge
of the first measurement outcome does not reveal any information about
the second measurement outcome. This we would call a fully uncorrelated
system. Let us quantify this statement via a correlation function
for two observables given in the spectral decomposition, $A_{k}$
and $B_{k}$ measured by Alice and Bob (with respective eigenvectors
$\left\{ |i_{k}\rangle\right\} =\left\{ |0_{k}\rangle,\dots,|d-1_{k}\rangle\right\} $)
by summing all \emph{joint probabilities} $P_{A_{k},B_{k}}(i_{k},i_{k})$
when both parties have the same outcome $i_{k}$
\begin{equation}
C_{A_{k},B_{k}}=\sum_{i=0}^{d-1}P_{A_{k},B_{k}}(i_{k},i_{k})=\sum_{i=0}^{d-1}Tr(|i_{k}i_{k}\rangle\langle i_{k}i_{k}|\;\rho_{d})\label{eq:jointprob}
\end{equation}
Here, we allow Alice and Bob to relabel their outcomes such that $C$
gets maximal. Indeed the above function is a \emph{mutual predictability},
since if the state is fully correlated then the outcomes can always
be labelled such that $C=1$ and if fully uncorrelated then any relabeling
can only give $C=1/d$ since all $d$ outcomes are equally likely.
However, this function does not tell us anything about entanglement,
since clearly the socks of Prof. Bertlmann \cite{bell1981}, who has
the habit of wearing differently coloured socks, are also fully correlated.
Therefore, in a second step, Alice and Bob will now use the fundamental
concept of complementarity and choose a second set of observables
that are mutually unbiased to the first choices of observables $A_{1}$
and $B_{1}$, namely the observables $A_{2}$ and $B_{2}$. One way
to phrase Bohr's complementarity of two observables $A_{1}$, $A_{2}$
is to say that they are non-degenerate; i.e., they do not share any
common eigenvector. This implies that the uncertainties of the outcomes
are bounded by the scalar product of the eigenvectors of both observables
and it is maximal if (and only if) all scalar products satisfy
\begin{equation}
|\langle i_{n}|j_{m}\rangle|^{2}=\frac{1}{d}\qquad\forall i,j\in{\{0,1,\dots,d-1\}}\;.\label{eq:mub}
\end{equation}

Thus, predictabilities of a product state $|0_{1}0_{1}\rangle$ are
given by
\begin{eqnarray}
P_{A_{1},B_{1}}(i_{1},i_{1}) & = & \delta_{i,0}\nonumber \\
P_{A_{2},B_{2}}(i_{2},i_{2}) & = & Tr\left(|i_{2}i_{2}\rangle\langle i_{2}i_{2}|0_{1}0_{1}\rangle\langle0_{1}0_{1}|\right)\nonumber \\
 & \stackrel{Eq.(\ref{eq:mub})}{=} & \underbrace{|\langle0_{1}|i_{2}\rangle|^{2}}_{\frac{1}{d}}\cdot\underbrace{|\langle0_{1}|i_{2}\rangle|^{2}}_{\frac{1}{d}}=\frac{1}{d^{2}},\label{eq:predict}
\end{eqnarray}
which holds for all $i$ and we obtain for the correlation functions
$C_{A_{1},B_{1}}=1$ and $C_{A_{2},B_{2}}=\frac{1}{d}$. For a third
choice of mutually complementary observables we obtain again $\frac{1}{d}$.
In general, by adding up $m$ correlation functions we obtain
\begin{equation}
I_{m}:=\sum_{k=1}^{m}C_{A_{k},B_{k}}\leq1+(m-1)\frac{1}{d}\label{eq:mubbound1}
\end{equation}
that has to hold for all pure separable states. In Ref.~\cite{spengler2012}
we have shown that it is also valid for all mixed separable states.
Therefore, $I_{k}$ serves as a detection function of entanglement
if and only if the bound is not satisfied. The existence of complete
sets of mutually unbiased bases in arbitrary dimensions is an open
problem, but in the case of prime-power dimensions $d$ it is known
to be $d+1$, leading to
\begin{equation}
I_{d+1}\stackrel{Eq.(\ref{eq:mubbound1})}{=}\sum_{k=1}^{d+1}C_{A_{k},B_{k}}\leq2\;.\label{eq:mubboundprime}
\end{equation}
It was shown that this detection criterion is very powerful as it
e.g. detects all the entanglement of isotropic states and more entanglement
of multiparticle Aharanov state than the powerful criteria of Refs.~{[}\onlinecite{hiesmayr2009,huber2010,gabriel2010}{]}.
Let us remark that our method of maximum complementarity provides
a test for non-separability of a state whereas Bell inequalities can
serve as an entanglement test but are designed to test for local realism.
Indeed, so far only multipartite bound entangled states were found
to violate a Bell inequality while the question is still open for
bipartite bound entangled states.

Now, we demonstrate both theoretically and experimentally that $I_{d+1}$
is also capable of detecting bound entanglement for a certain class
of so-called magic simplex states or Bell-diagonal states in $3\times3$
dimensions.

\begin{figure}
\includegraphics[width=1\columnwidth]{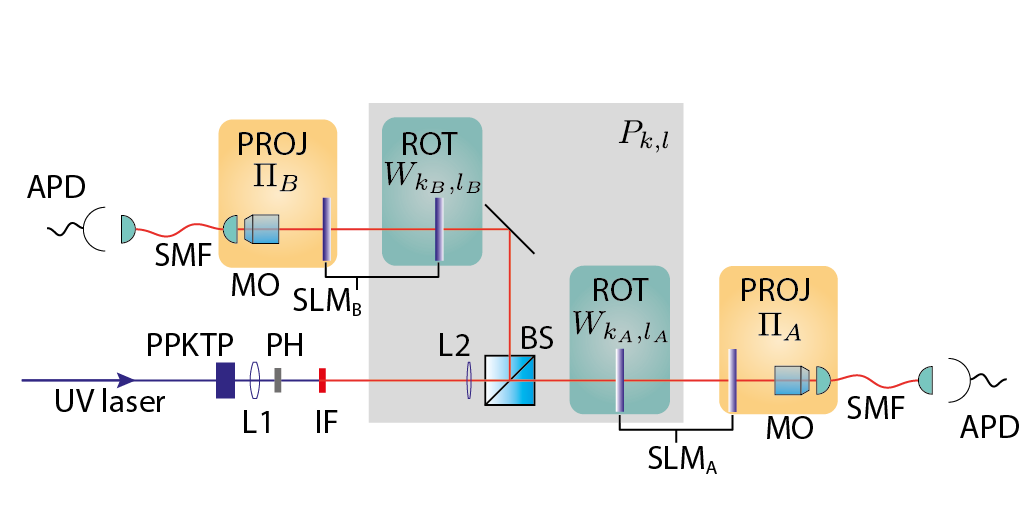}\caption{\label{fig:setup} Experimental setup for generation of bound-entangled
bipartite qutrits. Photon pairs are created by downconversion (type-I)
of 413~nm UV photons in a PPKTP crystal, momentum-filtered by pinhole
(PH), and split probabilistically using a beamsplitter (BS). Rotation
(ROT) in the orbital angular momentum superposition mode-space for
production of the Bell states $P_{k,l}$ (gray box) via $W_{k_{A/B},l_{A/B}}$
is performed using a spatial light modulator (SLM). Projective measurements
(PROJ) are done by appropriate mode transformation and subsequent
imaging onto the core of a single mode fibre (SMF) as indicated by
$\Pi_{A/B}$. We have implemented both operations on the same SLM.
The photons are guided to an avalanche photo diode (APD), and detection
events belonging to a single photon pair are post-selected by coincidence
detection.}
\end{figure}

For the experimental test, we have chosen to use orbital-angular-momentum
\cite{allen1992} (OAM) entangled photons generated by spontaneous
parametric downconversion (SPDC). OAM entanglement \cite{monken1998,mair2001,frankearnold2002,dada2011}
is one implementation of spatial entanglement which is particularly
intuitive because the quantum correlations of the photon pair are
determined only by conservation of orbital angular momentum during
pair generation: The OAM $\ell_{A}$ and $\ell_{B}$ of the downconverted
photons must fulfil $\ell_{A}+\ell_{B}=0$ if the pump is a flat (Gaussian)
beam with $\ell=0$ \cite{walborn2004,mair2001}. In the OAM basis
$|\ell_{A},\ell_{B}\rangle$ the two-photon qutrit state as produced
by the crystal is
\begin{eqnarray}
|\Psi_{SPDC}\rangle=\frac{1}{\sqrt{3}}\left\lbrace |-1,+1\rangle+|0,0\rangle+|+1,-1\rangle\right\rbrace .
\end{eqnarray}
In the following, we denote the states $\{|\ell=-1\rangle,|\ell=0\rangle,|\ell=+1\rangle\}$
by $\{|0\rangle,|1\rangle,|2\rangle\}$. As a starting point we take
the maximally entangled Bell state $P_{0,0}=|\Psi_{SPDC}\rangle\langle\Psi_{SPDC}|$
and, via applying the unitary Weyl operators $W_{k,l}:=\sum_{n=0}^{d-1}e^{\frac{(2\pi i)(kn)}{d}}|n\rangle\langle n+l|$
in one subsystem (e.g. on the photon of Alice), we can synthesize
the $d^{2}-1$ maximally entangled Bell states $P_{k,l}=W_{k,l}\otimes\mathbb{1}\, P_{0,0}\, W_{k,l}^{\dagger}\otimes\mathbb{1}$.
Any convex combination of these $d^{2}$ Bell states forms a so called
``\textit{magic simplex}'' \cite{baumgartner2006} $\mathcal{W}:=\{\rho_{d}=\sum_{k,l=0}^{d-1}c_{k,l}P_{k,l}|c_{k,l}\geq0,\sum_{l,k=0}^{d-1}c_{k,l}=1\}$.
This reduced state space of locally maximally mixed states admits
a simple geometrical representation (the vertices are the $d^{2}$
Bell states) and has been shown~\cite{baumgartner2006,baumgartner2007,baumgartner2008,bertlmann2009}
to be powerful in addressing inseparability issues and quantum information
theoretic questions.

For our purpose let us reduce the $d^{2}-1$ dimensional parameter
space to three (four in case of $d>3$) parameters $q_{i}$ and consider
the following family of states:
\begin{eqnarray}
\rho_{d} & = & (1-\frac{q_{1}}{d^{2}-(d+1)}-\frac{q_{2}}{d+1}-q_{3}-(d-3)q)\frac{1}{{d^{2}}}\mathbbm{1}_{d^{2}}\nonumber \\
 &  & +\frac{q_{1}}{d^{2}-(d+1)}P_{0,0}+\frac{q_{2}}{(d+1)(d-1)}\sum_{i=1}^{d-1}P_{i,0}\nonumber \\
 &  & +\frac{q_{3}}{d}\sum_{i=0}^{d-1}P_{i,1}+(1-\delta_{d,3})\frac{q_{4}}{d}\sum_{z=2}^{d-2}\sum_{i=0}^{d-1}P_{i,z}\label{eq:beatrixstate}
\end{eqnarray}
This family also includes for $d=3$ the one-parameter Horodecki--state,
the first found bound entangled state\cite{horodecki1998}. Namely,
for $q_{1}=\frac{30-5\lambda}{21},\; q_{2}=-\frac{8\lambda}{21},\; q_{3}=\frac{5-2\lambda}{7}$
with $\lambda\in[0,5]$. This state is PPT for $\lambda\in[1,4]$
and was shown to be bound entangled for $\lambda\in\{3,4]$.

\begin{figure}
\includegraphics[width=0.7\columnwidth]{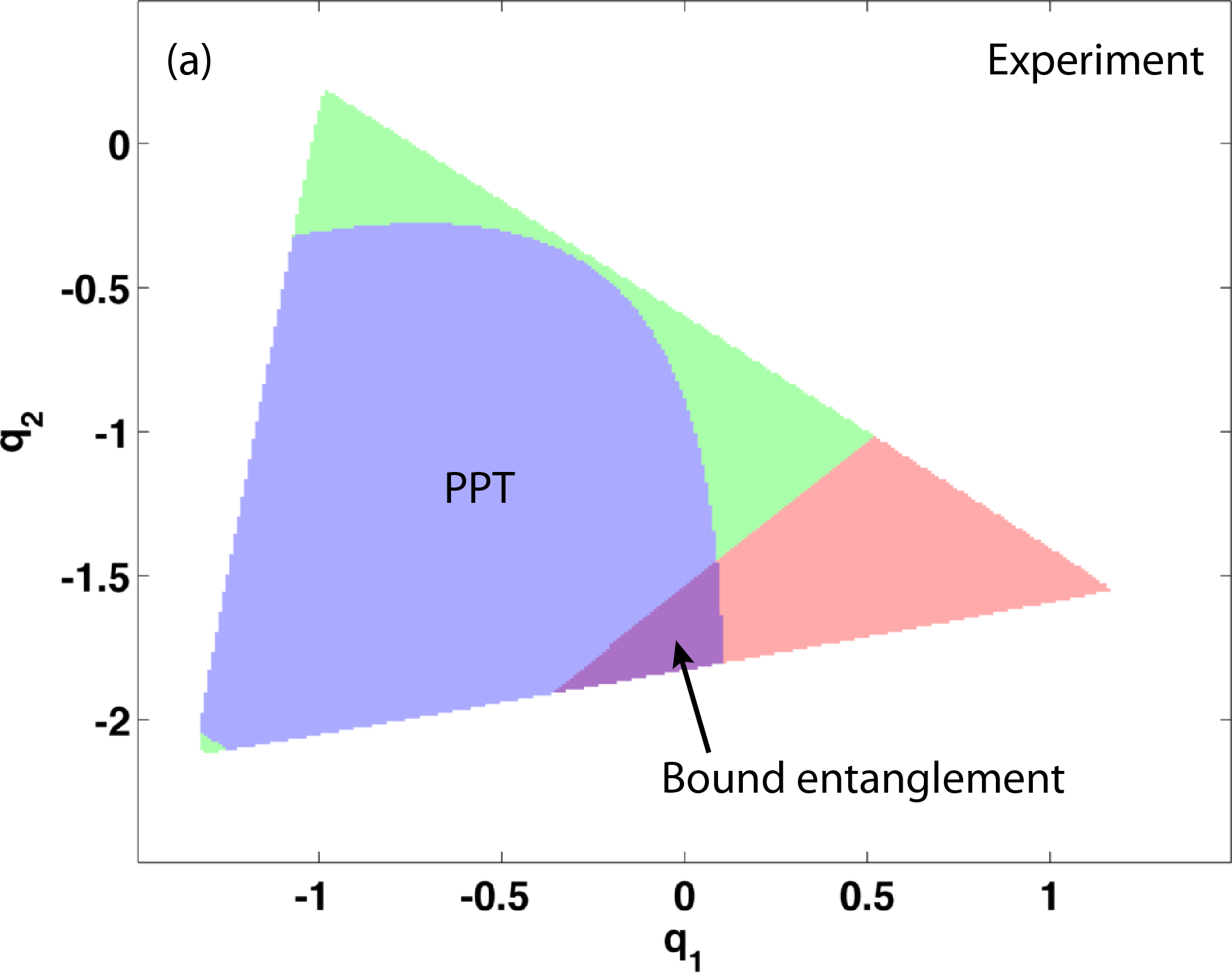} \includegraphics[width=0.7\columnwidth]{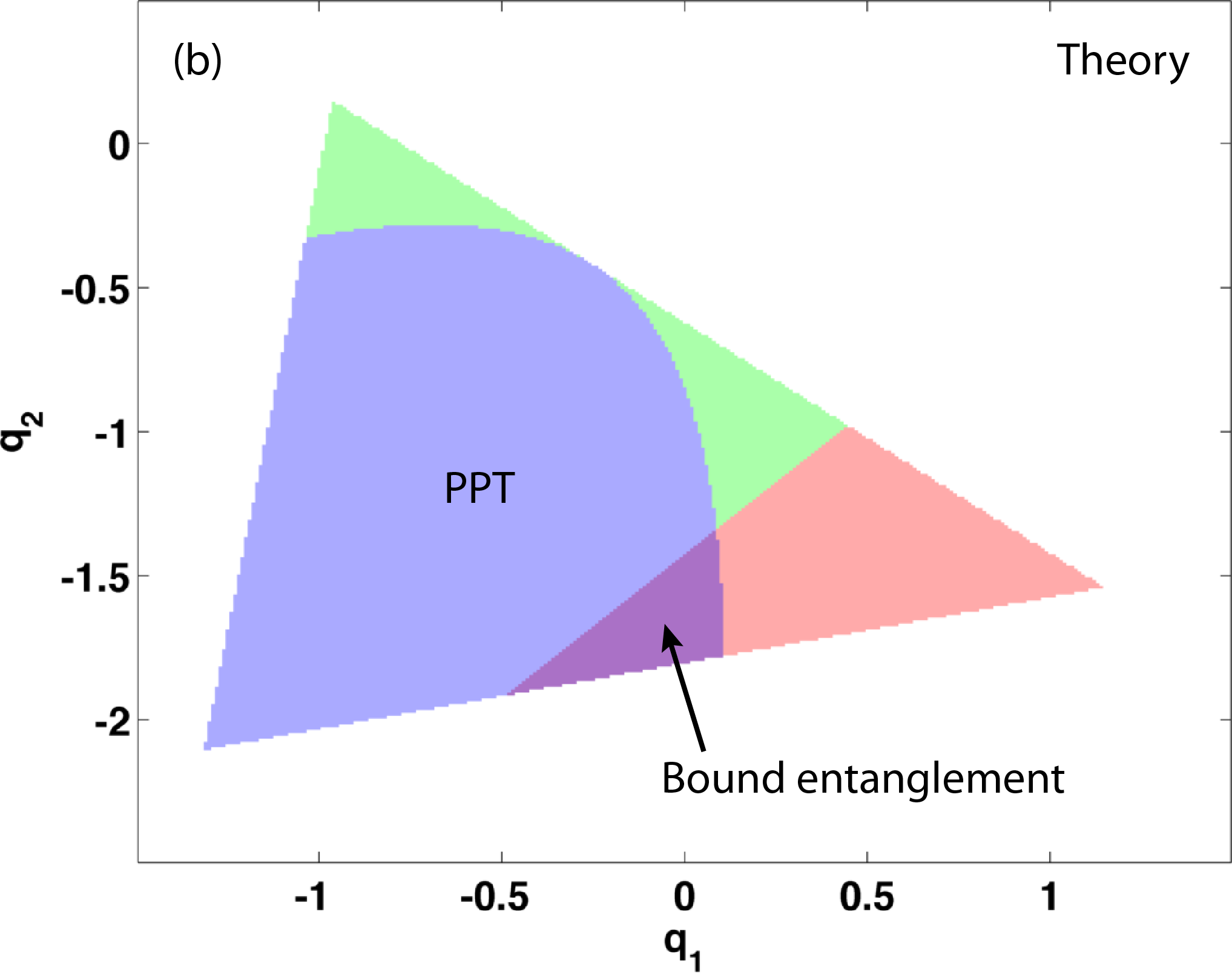}
\caption{\label{fig:qstresultsimplex2d} Experimental (a) and theoretical (b)
2D slice $\{q_{1},q_{2}\}$ through the magic simplex for fixed $q_{3}=-0.5776$.
All coloured points correspond to states having positive semidefinite
eigenvalues, therefore representing physical states. The blue area
(curved region) covers the range of states with a positive partial
transpose (PPT). The red triangular area indicates where the maximum
complementarity protocol applied to the experimentally generated states
(a) or theoretical states (b) is greater than $2$, thus detecting
entanglement. States where all three conditions are fulfilled are
bound entangled. We see that the experimental and theoretical geometries
agree very well. }
\end{figure}

\textbf{Generation of the states ($d=3$):} By expressing the totally
mixed state $\mathbb{1}_{9}$ as the sum of all Bell states $\sum P_{k,l}$
we find that all $9$ Bell states need to be mixed to synthesize the
state $\rho_{3}$. Note that there are $72$ unitary equivalent possibilities
\cite{baumgartner2006} to generate $\rho_{3}$ that we will exploit
to deduce the error (see supplementary information). We introduce
a measurement-based scheme to produce these mixed states similar to
the authors in Refs.~\cite{acin2009,amselem2009,lavoie2010b,dobek2011,kaneda2012}
for polarization but in our case for OAM qutrits using spatial light
modulation, see Fig.~\ref{fig:setup}. As described above we have
to apply single-qutrit rotations (e.g., on qutrit A) using the Weyl
matrix $W_{k,l}$ to transform $P_{0,0}$ into any of the $8$ other
maximally entangled Bell states $P_{k,l}$. This operation is implemented
on the spatial light modulators (SLMs, see Fig.~\ref{fig:setup}),
which allows us to generate the mixed state by time-multiplexing of
the rotation operators $W_{k,l}$ for a particular choice of $q_{i}$.
Our method is physically equivalent to tracing over a separate degree
of freedom to create decoherence, such as the spectral one, a method
which has also been used to generate mixed states \cite{peters2004,puentes2006}
or stochastic rotation with an additional optical element \cite{acin2009,amselem2009,lavoie2010b,dobek2011,kaneda2012}.
In our case we extend this to obtain full control over the high-dimensional
mixture by using \emph{retroactive mixing}: we record photon counts
for each of the states $P_{k,l}$, and form the incoherent summation
in a computer afterwards. This allows us to explore the magic simplex
${\mathcal{W}}$ completely, while still obtaining exactly the same
result as if mixing would happen during the photon counting time.
Projective measurements are performed by mode-conversion on the spatial
light modulator and imaging onto a single-mode fibre.

\begin{figure}
\includegraphics[width=0.7\columnwidth]{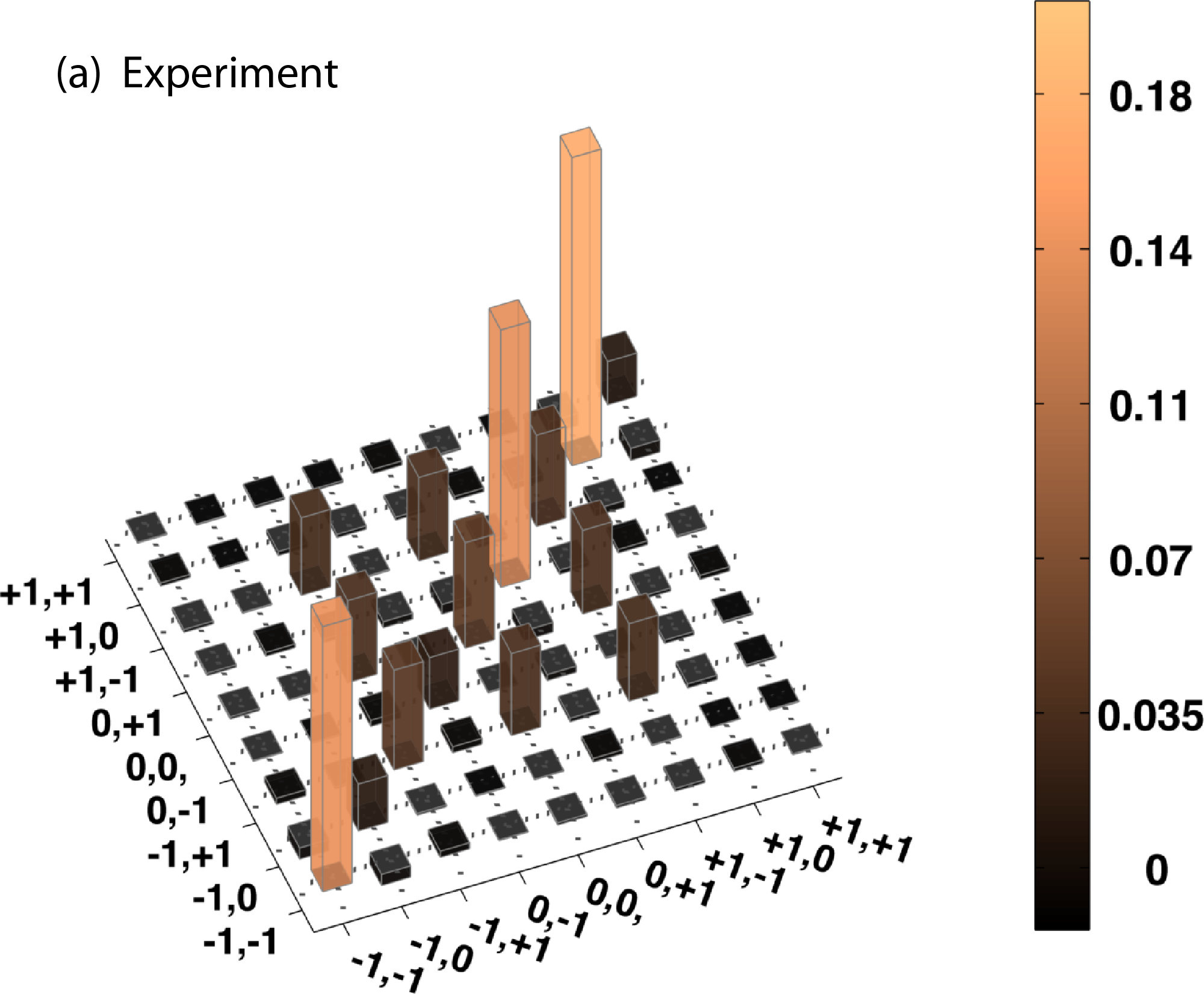} \includegraphics[width=0.7\columnwidth]{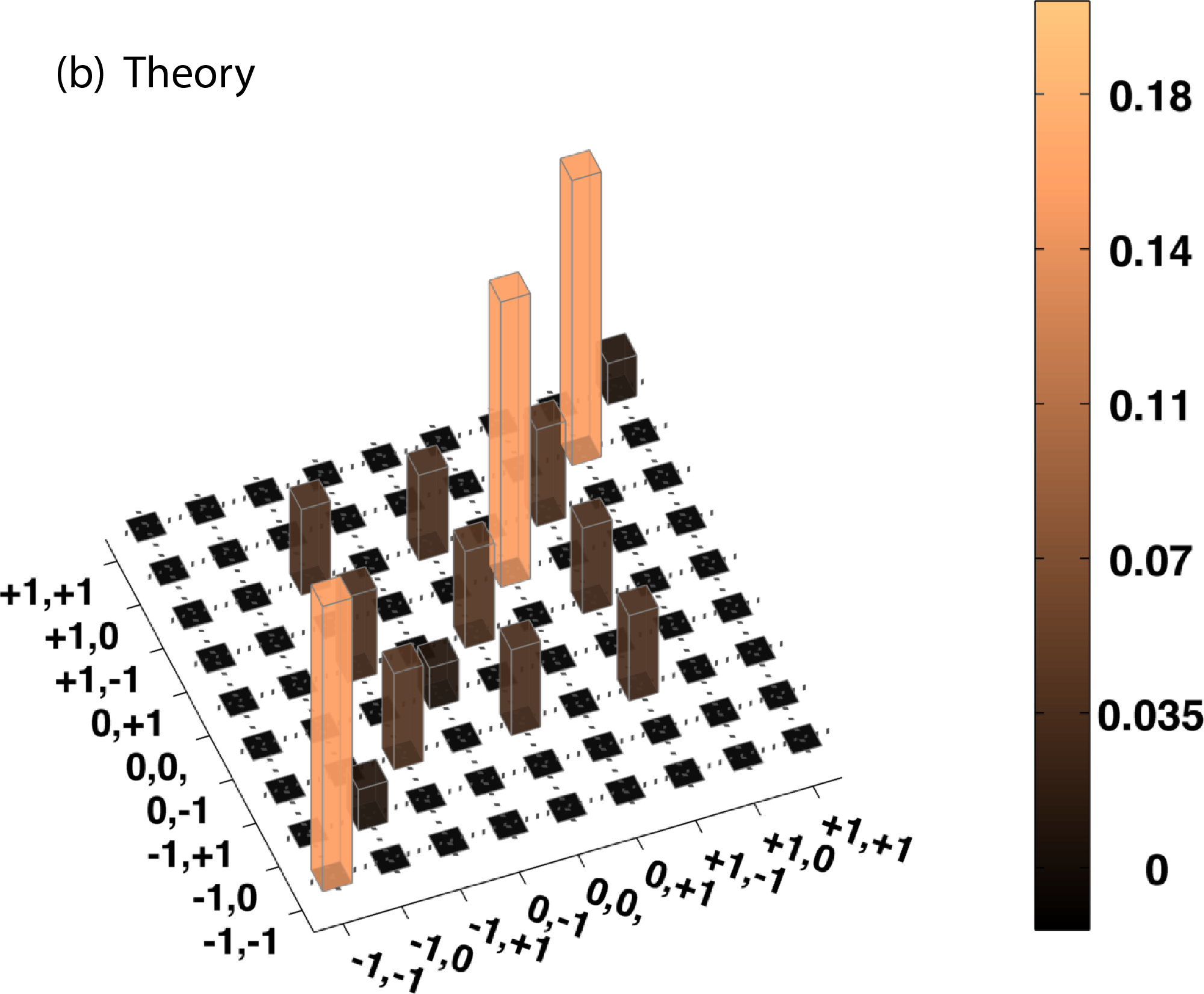}
\caption{\label{fig:qstresultrhos} Experimental (a) and theoretical (b) real
parts of the density matrices of the bound entangled state ($q_{1}=-0.07,\, q_{2}=-1.73,\, q_{3}=-0.5774$).
The fidelity of the experimental state is 0.98, the complementarity
protocol applied to the experimental state gives $2-I_{d+1}=-0.032\pm0.003$
and the minimum eigenvalue of the partially transposed state is $\textrm{Min}\left[\textrm{eig}(\rho^{T_{A}})\right]=+0.0101\pm0.0017$;
the state is bound entangled. It is expected to be purely real; the
experimental imaginary parts are all below 10\%. The axes labels indicate
the OAM quantum numbers $\ell$.}
\end{figure}

\textbf{State Tomography and Magic Simplex:} We first perform quantum
state tomography \cite{james2001} of the bound entangled state. We
keep $q_{3}=-0.5776$ fixed, this allows us to visualize quantum state
properties in 2D: Fig.~\ref{fig:qstresultsimplex2d} shows experimental
(a) and theoretical case (b), where each coloured pixel $\{q_{1},q_{2}\}$
corresponds to a state, blue if PPT (curved region), and red if the
state violates the maximum complementarity protocol, $2-I_{d+1}<0$
(triangle). In regions where all conditions are fulfilled, we have
non-separable states with a positive partial transposition, i.e. bound
entanglement. We find that, compared to the total area of physical
states, the bound entangled states occupy a significant space in the
chosen slice through the magic simplex; the extension along $q_{3}$
turns out to be even much larger, i.e., the bound entangled states
occupies a rather large volume. Since the regions in Fig.~\ref{fig:qstresultsimplex2d}
are continuous (``they do not have holes''), it is obvious from
the geometry that states within the region where all criteria apply
are bound entangled. However, it is important to check that the conditions
are also fulfilled significantly \cite{lavoie2010}. In contrast to
the polarization qubit case \cite{amselem2009}, in spatial entanglement,
the most important errors are due to unwanted rotations of the state
due to wavefront aberrations. To quantify these errors, we have determined
for all $72$ unitary-equivalent states of $\rho_{3}$ (see supplementary
information) the result of the complementarity protocol $2-I_{d+1}$
and the smallest eigenvalue of the partially transposed state $\textrm{Min}\left[\textrm{eig}(\rho^{T_{A}})\right]$.
This allows us to derive realistic error estimates. In Fig.~\ref{fig:qstresultrhos}
we plot explicitly the matrix elements of the bound entangled state
($q_{1}=-0.07,\, q_{2}=-1.73,\, q_{3}=-0.5774$) together with the
theoretical prediction. The numerical values are $2-I_{d+1}=-0.032\pm0.003$
and $\textrm{Min}\left[\textrm{eig}(\rho^{T_{A}})\right]=+0.0101\pm0.0017$.
This clearly proves that the state is bound entangled: non-separable,
but PPT.

We studied also the historical Horodecki--state \cite{horodecki1998},
and we can indeed experimentally confirm bound entanglement thereof
for a large range of the $\lambda$-parameter. For instance, for $\lambda=3.5$,
we get $2-I_{d+1}=-0.025\pm0.003$, and $\textrm{Min}\left[\textrm{eig}(\rho^{T_{A}})\right]=+0.012\pm0.002$.

\bigskip

\textbf{Maximum Complementarity Protocol:} In a next step we want
to verify entanglement of the state, shown in Fig.~\ref{fig:qstresultrhos},
directly by using the complementarity protocol. For that, the correlation
functions had to be measured directly, which is considered to be difficult
due to normalization: In quantum state tomography, proper normalization
is imposed during numerical search; here we have to obtain probabilities
directly from experimental coincidence rates. We obtain the probabilities
$P$ from the normalized coincidence rates $\Gamma$ via $P_{A_{k},B_{k}}(i_{k},i_{k})=\Gamma_{A_{k},B_{k}}(i_{k},i_{k})/\sum_{s_{k},t_{k}}\Gamma_{A_{k},B_{k}}(s_{k},t_{k})$.
From this we calculate the correlation functions $C$ and obtain:\\
 \global\long\def\arraystretch{1.4}
\\
\begin{tabular}{|l|c|c|}
\hline
Correlation function  & Theory  & Experiment\tabularnewline
\hline
\hline
$C_{A_{1},B_{1}}$  & $0.675$  & $0.667\pm0.005$ \tabularnewline
\hline
$C_{A_{2},B_{2}}$  & $0.468$  & $0.463\pm0.005$\tabularnewline
\hline
$C_{A_{3},B_{3}}$  & $0.468$  & $0.469\pm0.005$\tabularnewline
\hline
$C_{A_{4},B_{4}}$  & $0.468$  & $0.467\pm0.005$\tabularnewline
\hline
\hline
$2-\left(I_{4}=\sum_{k}C_{A_{k},B_{k}}\right)$  & $-0.079$  & $-0.066\pm0.02$\tabularnewline
\hline
\end{tabular}

The uncertainties are determined from multiple experimental runs.
Experiment and theory agree well, the maximum complementary protocol
confirms that we detect a truly entangled state. In literature, criteria
for entanglement are usually applied on density matrices (as we did
in the first step). We have performed extensive experimental tests
and conclude that the maximum complementarity protocol can directly
be measured if the experiment is well under control. This implies
a significant advance in the exploration of high-dimensional entangled
photons, since the experimental and computational complexity is strongly
reduced: We need $N_{QST}=d^{2}-4d^{3}+4d^{4}$ measurements for (over-complete)
state tomography, and only $N_{MCP}^{(1)}=d+d^{2}$ measurements to
determine the function $I_{d+1}$ directly (without normalization),
and $N_{MCP}^{(2)}=d^{2}+d^{3}$ including normalization.

\textbf{Generality of the Maximum Complementarity Protocol:} Let us
discuss what we would expect for the case of higher dimensional OAM
entangled photons in the state $\rho_{d}$. We search the parameter
space of prime and prime-power dimensional entangled states (the dimensions
where $d+1$ MUBs are known), and optimize our maximum complementarity
protocol to find the states where the detection of entanglement via
$2-I_{d+1}$ is strongest:
\begin{eqnarray}
\min_{q_{i},\rho_{d}\geq0,\rho_{d}^{T_{A}}\geq0}2-I_{d+1}[\rho_{d}]=\left\lbrace \begin{array}{c}
-0.15\,(d=3)\\
-0.125\,(d=4)\\
-0.106\,(d=5)\\
-0.081\,(d=7)\\
-0.073\,(d=8)\\
-0.067\,(d=9)
\end{array}\right.
\end{eqnarray}
In each case, a similar geometry as shown in Fig.~\ref{fig:qstresultsimplex2d}
is obtained. In dimension $d=2\cdot3=6$ so far only $3$ MUBs are
found and there are strong numerical hints \cite{durt2010} that there
exist not more; applying only these does not lead to the detection
of bound entanglement. In Ref.~\cite{hiesmayr2009b} we have shown
via geometry considerations that if a Hermitian operator detects entanglement
of states in a certain quantum space $\mathcal{W}$, then it also
detects entanglement in the multi-partite product space $\mathcal{W}^{\otimes n}:=\{\rho_{d}^{\otimes n}=\sum c_{k,l}\tilde{P}_{k,l}|c_{k,l}\geq0,\sum c_{k,l}=1\}$.
Also in this case, the $d^{2}$ Bell-type vertex states $\tilde{P}_{k,l}$
are obtained by applying a Weyl operator in one subsystem to $\tilde{P}_{0,0}=\frac{1}{d^{2}}\sum P_{k,l}^{\otimes n}$.
Therefore, the maximum complementarity protocol is also applicable
for multipartite states and can detect bound entanglement therein.

\noindent To illustrate the richness and difference between bipartite
and multipartite entanglement, let us discuss first the famous case
of $d=2$. Then the vertex states are the Smolin states~\cite{smolin2001}
that are known to be multiparticle unlockable bound entangled, because
no local party can distil entanglement; however, \emph{two} parties
can distil (unlock) the entanglement, which would in this case be
actually a pure maximally entangled state; but it is only available
for the other parties, not themselves. This case was recently experimentally
demonstrated \cite{amselem2009,lavoie2010b}. Surprisingly, entangled
states inside the magic simplex $\mathcal{W}^{\otimes n}$ can be
distilled to the Smolin-type vertex states \cite{hiesmayr2009b},
which themselves are undistillable. For $d>2$, in addition, the multipartite
magic simplex contains $PPT$-entangled states, therefore, states
that cannot even be purified to a vertex state. They are bound entangled
and only ``bound unlockable'' due to their multiparticle nature.

We have proven experimentally that bipartite bound entanglement exists
in Nature, which was predicted in 1998~\cite{horodecki1998} and
started an intensive theoretical quest in understanding its meaning.
We have chosen orbital angular momentum entanglement of photons that
is scalable in the dimensionality \cite{dada2011}, and our high-quality
results suggest that further exploration of high dimensional entangled
quantum states is possible. Therefore, our experimental method and
the maximum complementarity protocol is a useful addition to the toolbox
to explore different types of entanglement including bound entanglement.
With the case of two qutrits as the most simple system we lay the
foundation to pursue the question why Nature should provide us with
such a strange form of highly mixed entanglement that can not be purified,
despite the fact we have used pure maximally entangled states as a
resource to produce the state: What is its physical meaning? What
are the quantum information theoretic applications of high-dimensional
bound entanglement?

\noindent Finally, via the maximum complementarity protocol we provide
an alternative proof that the maximum number of mutually unbiased
bases (MUBs) cannot be more than $d+1$: If for families of states
that are optimally detected by $I_{d+1}$, another MUB would be added,
we would detect separable states as entangled ones. The minimum number
of existing MUBs is known to be the smallest of the prime factors
of $d+1$. It is also open whether one always needs all MUBs to detect
bound entanglement. Consequently, investigating inseparability problems
opens a different trail to look for a solution of the number problem
of MUBs.

\textbf{Methods:}

\emph{Maximum complementarity protocol}. As an explicit example we
show our choice of basis vectors of the four mutually unbiased bases
$\mathcal{B}_{k}$ with $w=e^{\frac{2\pi i}{3}}$:
\begin{eqnarray*}
\mathcal{B}_{1} & : & \left\{ |0_{1}\rangle,|1_{1}\rangle,|2_{1}\rangle\right\} =\left\{ \left(\begin{array}{c}
1\\
0\\
0
\end{array}\right),\left(\begin{array}{c}
0\\
1\\
0
\end{array}\right),\left(\begin{array}{c}
0\\
0\\
1
\end{array}\right)\right\} \\
\mathcal{B}_{2} & : & \left\{ |0_{2}\rangle,|1_{2}\rangle,|2_{2}\rangle\right\} =\left\{ \left(\begin{array}{c}
1\\
1\\
1
\end{array}\right),\left(\begin{array}{c}
1\\
w\\
w^{2}
\end{array}\right),\left(\begin{array}{c}
1\\
w^{2}\\
w
\end{array}\right)\right\} \\
\mathcal{B}_{3} & : & \left\{ |0_{3}\rangle,|1_{3}\rangle,|2_{3}\rangle\right\} =\left\{ \left(\begin{array}{c}
1\\
w\\
w
\end{array}\right),\left(\begin{array}{c}
1\\
w^{2}\\
1
\end{array}\right),\left(\begin{array}{c}
1\\
1\\
w^{2}
\end{array}\right)\right\} \\
\mathcal{B}_{4} & : & \left\{ |0_{4}\rangle,|1_{4}\rangle,|2_{4}\rangle\right\} =\left\{ \left(\begin{array}{c}
1\\
w^{2}\\
w^{2}
\end{array}\right),\left(\begin{array}{c}
1\\
1\\
w
\end{array}\right),\left(\begin{array}{c}
1\\
w\\
1
\end{array}\right)\right\}
\end{eqnarray*}
To detect (bound) entanglement, we choose the following combination
of correlation functions ($|i^{\ast}\rangle=|i\rangle^{\ast}$):
\begin{eqnarray*}
C_{A_{1},B_{1}} & = & \sum_{i=0}^{2}Tr(|i_{1},\textrm{mod}(i_{1}+1,3)^{*}\rangle\langle i_{1},\textrm{mod}(i_{1}+1,3)^{*}|\rho_{3})\\
C_{A_{k},B_{k}} & = & \sum_{i=0}^{2}Tr(|i_{k},i_{k}^{*}\rangle\langle i_{k},i_{k}^{*}|\rho_{3}),\; k=2,3,4\;.
\end{eqnarray*}

\emph{Experiment.} We generate the spatially entangled photon pairs
by collinear Type-I SPDC in a PPKTP crystal (length $L$~=~2~mm)
of a $LG_{0}^{0}$ laser beam ($\mathsf{Kr^{+}}$, $\lambda$~=~413~nm,
beam waist at crystal $w_{p}$~=~325~\textgreek{m}m, 80~mW power).
The temperature is tuned to detect a similar amount of downconverted
photons in the $\ell=0$ and $\ell=\pm1$ OAM mode. We image the crystal
surface with $7.5\times$ magnification using a telescope onto the
SLM (Hamamatsu X10468-07) surface. The SLM is operated under an incident
angle of 10 or 5 degrees; this allows us to use a single SLM for both
signal and idler photon. The (phase-corrected) SLM is used with a
blaze angle of 1 mrad. The elimination of the zeroth order is done
in a time-reversed fashion \emph{before} the SLM with a pinhole in
the far field of the crystal behind L1 (Fig.~\ref{fig:setup}). The
far field of the SLM surface is sent to the single mode fibre using
$10\times$ objectives, with a detection-fibre mode waist at the SLM
of 1275~\textgreek{m}m. The fibres are connected to single-photon
counters and we post-select photon pairs by coincidence detection
(time window 7~ns). All measurements were integrated for 2~s; we
obtain typically 30,000 single counts and 1,500 coincidence counts
for conjugated-field detector settings and <10 coincidence counts
for the cases where no coincidences are expected. The SLM kinoforms
required for 2-state superpositions involve only phase modulation,
however, for 3-state superpositions such as those used for implementation
of the maximum complementarity protocol, require spatial amplitude
modulation, too. Because this is very inefficient if done in a continuous
way \cite{dada2011}, we employ binary amplitude modulation by removing
the blaze at spatial positions where the normalized modulus of the
detection field amplitude is smaller $\sqrt{0.5}$. For tomographic
reconstruction, we obtain an overcomplete set of measurements by detecting
all two-state superpositions with relative phases of $\{0,\pi/2,\pi,3\pi/2\}$.

\textbf{Supplementary information:}

\begin{figure}
\includegraphics[width=0.8\columnwidth]{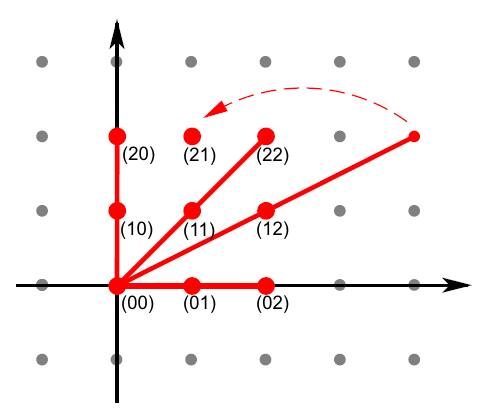} \caption{\label{fig:lines} Illustration of the finite discrete classical phase
space for dimension $d=3$ of the locally maximally mixed states of
the magic simplex $\mathcal{W}$. Each point $(kl)$ represent one
of the nine Bell states $P_{k,l}$. All possible complete lines through
the point $(00)$ for $d=3$ are drawn, representing one class of
states which have the same geometry concerning separability, bound
entanglement, free entanglement and nonlocality, i.e., those that
are unitary equivalent. The same holds for each line which is parallel
to any red line.}
\end{figure}

\emph{Phase space of two qutrits.}\textbf{ }Below, in Fig.~\ref{fig:lines},
the finite phase space of the magic simplex~\cite{baumgartner2006}
is drawn for dimension $d=3$. This space is spanned by all nine Bell
states $P_{k,l}$, which can be generated by applying the Weyl operators
$W_{k,l}$ onto one arbitrary maximally entangled Bell state, denoted
by $P_{0,0}$: $P_{k,l}=W_{k,l}\otimes\mathbb{1}\, P_{0,0}\, W_{k,l}^{\dagger}\otimes\mathbb{1}$.
Due to the group structure of the Weyl operators certain mixtures
of the Bell states $P_{k,l}$ are geometrically equivalent, i.e. have
the same properties concerning separability, bound entanglement, free
entanglement and nonlocality (violation of a given Bell inequality).
Thus one has not to analyze all possible mixtures since some are equivalent
concerning the properties we are interested in. In particular, the
so called \emph{lines} are special. A line is formed by choosing one
Bell state, e.g. $P_{0,0}$, and applying to it $d-1$ times the same
Weyl operator, e.g. $W_{0,1}$. Applying another time the Weyl operator
brings one back to to original state (in our case $P_{0,0}$), because
of the periodicity of the Weyl operators. In our case the three Bell
states $\{P_{0,0},P_{0,1},P_{0,2}\}$ form a line in the phase space
(see Fig.~\ref{fig:lines}). We find $d+1$ lines with different
orientations through one Bell state $P_{0,0}$ (red lines in Fig.~\ref{fig:lines}).
To each (red) line there are $3$ parallel lines. In summary, we find
$(3+1)\times3=12$ lines, which have the same geometry regarding separability,
bound entanglement, free entanglement, and nonlocality.

For our state $\rho_{3}$ (Eq.~7), we see that the states weighted
by $q_{1}$ and $q_{2}$ form a line and $q_{3}$ is the coefficient
for another line that is parallel to the first one. Thus, we have
$(3+1)\times3$ possibilities to choose a line and $3$ possibilities
to weight them with $q_{1}$ and $q_{2}$. And there are $2$ possibilities
to choose parallel lines to the chosen one weighted by $q_{3}$. Thus
we have in total $(3+1)\times3\times3\times2=72$ unitary equivalent
possibilities to obtain the state $\rho_{3}$, and all of these have
in theory the same geometry regarding separability, bound entanglement,
free entanglement and nonlocality.

In the experiment, the nine Bell states are not fully unitary equivalent
due to wavefront errors, thus we get for all $72$ possibilities slightly
different values that allows to perform statistical analysis to obtain
the true experimental errors for the complementarity protocol $2-I_{d+1}$
and the minimum eigenvalue of the partial transpose $\textrm{Min}\left[\textrm{eig}(\rho^{T_{A}})\right]$.

\textbf{Acknowledgements}

We thank R.A. Bertlmann and A. Gabriel for discussions and acknowledge
the SoMoPro programme, NWO, Austrian Science Fund (FWF-P23627-N16),
and the EU STREP program 255914 (PHORBITECH). The Project is funded
from the SoMoPro programme. Research of B.C.H. leading to these results
has received a financial contribution from the European Community
within the Seventh Framework Programme (FP/2007-2013) under Grant
Agreement No. 229603. The research is also co-financed by the South
Moravian Region. W.L. acknowledges support from NWO and the EU STREP
program 255914 (PHORBITECH).

\bibliographystyle{naturemagw}
\bibliography{bibliography}

\end{document}